\documentclass{appolb}
\usepackage[dvips]{graphicx}

\def\ni{\noindent}

\title{
\centerline{\bf Primordial clouds under the influence of background radiation field}
}

\author{
J. Stasielak$^{a}$
\and
S. Stachniewicz$^{b}$
\and
M.  Kutschera$^{a,b}$
\address{
$^a$Institute of Physics, Jagiellonian University, ul. Reymonta 4,\\
30-059 Krak\'ow, Poland \\
$^b$H.Niewodnicza\'nski Institute of Nuclear Physics, ul. Radzikowskiego 152,\\
31-342 Krak\'ow, Poland
}}

\begin{document}
\maketitle

\begin{abstract}
Our goal is to study the effects of the UV radiation from the first stars, quasars and hypothetical Super Heavy Dark Matter (SHDM) particle decays on the formation of primordial bound objects in the Universe. We trace the evolution of a spherically symmetric density perturbation in the Lambda Cold Dark Matter and MOND model, solving the frequency-dependent radiative transfer equation, non-equilibrium chemistry, and one-dimensional gas hydrodynamics. We concentrate on the destruction and formation processes of the $H_{2}$ molecule, which is the main coolant in the primordial objects. 
\end{abstract}


\section{Introduction}

Formation of primordial objects such as young galaxies and globular clusters, in the early universe is a fundamental problem in the modern cosmology. There was a rapid progress of observations in the last years. It gave theoreticians invaluable possibility to compare their theories with the observations. Now, we can say that the framework of the structure formation is well known. However, there are still many unanswered questions, especially regarding the influence of the background radiation field from the first objects to the star formation rate. In our work we are trying to answer this question. 

The existence of the first objects is a direct consequence of the growth of the primordial density fluctuations. At the beginning, there are linear density perturbations which expand with the overall Hubble flow. Subsequently, these perturbations can grow and form primordial clouds. Clouds with enough density contrast decouple from this flow and start to collapse. The kinetic energy of the infalling gas is dissipated through shocks and the cloud becomes pressure supported. The further evolution of the cloud is determined by its ability to cool sufficiently fast. Clouds which could not cool fast enough will stay in a pressure-supported stage and will not form any stars. The existence of the efficient cooling mechanism is necessary to continue the collapse of the cloud, its subsequent fragmentation and star formation. 

In our work we are interested in the first generation of stars, say III population, which are still forming in the low mass clouds when first luminous objects already exist. These objects are made from the primordial gas so they are metals free. It is simply because the first stars did not have much time to produce them. These objects could be irradiated by the UV and X-rays radiation produced by the first stars, quasars and hypothetical Super Heavy Dark Matter (SHDM) particle decays (\cite{Do02}, \cite{Sh04}).

In the absence of metals, the most important cooling mechanism for low-mass primordial clouds is so called '$H_{2}$ cooling', i.e. cooling by radiation of excited rotational and vibrational states of $H_{2}$ molecule. We can look at this mechanism as the collisional excitation of the $H_{2}$ molecule, subsequent spontaneous de-excitation and photon emission. Emitted photon can further escape from the cloud and take away kinetic energy of the collapsing cloud. The presence of initial mass fraction of the molecular hydrogen $H_{2}$ of only $10^{-6}$ is enough to trigger the final collapse of low mass clouds.

Molecular hydrogen is fragile and can easily be photodissociated by photons with energies of 11.26 - 13.6 eV (Lyman and Werner bands)(\cite{Ms82}) through the following reactions
\[
	H_{2}+h\nu \rightarrow H^{*}_{2}\rightarrow 2H \textrm{.}
\]
Destruction of the $H_{2}$ would stop collapsing of the low mass clouds and decrease star formation rate.

In primordial gas cloud, $H_{2}$ molecules can form mainly through the coupled reactions
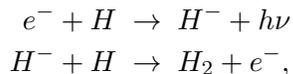
\begin{eqnarray}
	e^{-} + H& \rightarrow &H^{-} + h\nu \nonumber \\
	H^{-}+H & \rightarrow & H_{2} + e^{-} \nonumber \textrm{,}
\end{eqnarray}
in which the electrons act only as a catalyst. X-rays radiation can increase production of the $H_{2}$ by enhancement of free electrons fraction.

From above we see that the UV and X-rays radiation background alters the subsequent growth of cosmic structures. It regulates star formation rate so it has important implication to the re-ionization history of the Universe \mbox{(\cite{Ce03})}. In addition photoionization by UV photons can be regarded as the process which lead to the inhibition of the low mass galaxies formation, so it could explain the so called 'low-mass galaxies overproduction problem' in the hierarchical bottom-up theory of galaxy formation. It is therefore crucial to determine quantitatively the consequences of radiation feedback on the formation of early generation objects.

Feedback of the UV background to the collapse of the spherically symmetric primordial gas cloud in CDM model was studied by many authors (\cite{Ta98}, \cite{Ki00}, \cite{Ki00b}, \cite{Ki00c}, \cite{Om01}). However, their calculations were simplified. They have been using self-shielding function (\cite{Dr96}) instead of solving the full transfer equation. Our approach includes solving the frequency-dependent radiative transfer equation in the 'exact' way. We try to check the evolution of the collapsing cloud in the $\Lambda$CDM and MOND model.

\section{Basic equations}

In Cold Dark Matter models and in case of spherical symmetry, the collapse 
may be described by the following equations:

\begin{eqnarray}
{dM \over dr} & = & 4\pi r^2 \varrho, \label{ciaglosc}\\
{dr \over dt} & = & v , \label{promien}\\
{dv \over dt} & = & -4\pi r^2 {dp \over dM}-{GM(r) \over r^2} ,
\label{predkosc}\\
{du \over dt} & = & {p \over \varrho^2} {d \varrho \over dt} + {\Lambda \over
\varrho} ,
\label{energia}
\end{eqnarray}

\ni where $r$ is the radius of a sphere of mass $M$, $u$ is the internal
energy per unit mass, $p$ is pressure and $\varrho$ is mass
density. Here, Eq.(\ref{ciaglosc}) is the continuity equation, (\ref{promien})
and (\ref{predkosc}) give acceleration and (\ref{energia}) accounts for
energy conservation.
The last term in Eq.(\ref{energia}) describes gas cooling/heating, where
$\Lambda$ is energy absorption (emission) rate per
unit volume.

$\Lambda$ consist of two parts, that is, the chemical cooling $\Lambda _{chem}$ and the radiative cooling $\Lambda_{rad}$:
\begin{equation}
	\Lambda\left(r\right)=\Lambda_{chem}\left(r\right)+\Lambda_{rad}\left(r\right) \textrm{.}
\end{equation}
The $\Lambda_{chem}$ can be written as
\begin{equation}
\Lambda_{chem} = \varrho \frac{\partial \epsilon _{chem}}{\partial t}
\end{equation}
\noindent where $\epsilon _{chem}$ is the chemical binding energy per unit mass. 
The radiative cooling $\Lambda_{rad}\left(r\right)$ is obtained by solving radiative transfer equation.

We use the equation of state of perfect gas

\begin{equation} p= (\gamma -1) \varrho u , \end{equation}

\ni where $\gamma = 5/3$, as the primordial baryonic matter after
recombination
is assumed to be composed mainly of monoatomic hydrogen and helium, with
the fraction of molecular hydrogen $H_2$ always less than $10^{-3}$.

In the case of modified gravity\cite{Mil83}, equation (\ref{predkosc}) 
will change:

\begin{equation}
{dv \over dt}= -4\pi r^2 {dp \over dM}-g_H-a_0f\left({GM(r) \over a_0 r^2}-{g_H \over a_0}\right) ,
\label{modpr}
\end{equation}

\ni where $f(x)$ is some function, asymptotically equal to $x$ for $x\gg a_0$ 
and to $\sqrt{a_0x}$ for $x\ll a_0$, while $g_H$ may be expressed as

\begin{equation}
g_H = {1 \over 2} {H_0}^2 \left[(z+1)^3\Omega_b + 2 \left( (z+1)^4\Omega_r -
\Omega_{\Lambda}\right) \right] r .
\label{gH}
\end{equation}

\ni Here, $H_0$ is the current value of the Hubble parameter, $\Omega_b$,
$\Omega_r$ and $\Omega_{\Lambda}$ are the current fractions
of baryons, radiation and dark energy in terms of the critical density of the
Universe, and $z$ is the redshift. We have made similar assumption like in 
\cite{Sta05} and we apply MOND to net acceleration over Hubble flow only.

\section{Radiative transfer equation in spherical symmetry}

The time-independent, non-relativistic equation for radiation transport in spherical geometry can be written:
\begin{equation}
\mu \frac{\partial I_{\nu}}{\partial r} + \frac{1 - \mu^{2}}{r} \frac{\partial I_{\nu}}{\partial \mu} =
\varrho \left\{ \eta_{\nu}\left(r, \mu \right)	-  \chi_{\nu}\left(r\right) I_{\nu}		\right\} \label{sfer} \textrm{,}
\end{equation}
where $I_{\nu} = I_{\nu} \left(r,\mu\right)$ is the intensity of radiation of frequency $\nu$, at radius $r$ and in the direction $\mu = \cos \theta$, where $\theta$ is the angle between the outward normal and photon direction. $\eta_{\nu}\left(r, \mu\right)$ and $\chi_{\nu}\left(r\right)$ are the total emissivity and opacity at frequency $\nu$.

$\eta_{\nu}\left(r, \mu\right)$ and $\chi_{\nu}\left(r\right)$ can be expressed as:
\begin{eqnarray}
	\eta_{\nu}\left(r, p\right) &=& \eta^{t}_{\nu}\left(r\right) + \eta^{s}_{\nu}\left(r, p\right) \\
	\chi_{\nu}\left(r\right) &=& \kappa_{\nu}\left(r\right) + \sigma_{\nu}\left(r\right) \textrm{,}
\end{eqnarray}
where $\eta^{t}_{\nu}$, $\eta^{s}_{\nu}$, $\kappa_{\nu}$ and $\sigma_{\nu}$ are thermal emissivity, scattering emissivity, true absorption coefficient and scattering coefficient at frequency $\nu$ respectively.

Following Hummer and Rybicki \cite{hu71}, we introduce a more convenient system of coordinates $\left(r, p\right)$ rather than $\left(r, \mu\right)$ defined by the transformation formulae
\begin{equation}
\left(r, \mu \right) \rightarrow \left(r, p = r\sqrt{1-\mu^{2}}\right) \label{variable} \hskip 5mm
\textrm{for} \hskip 5mm -1\leq \mu \leq 1 \textrm{.} 
\end{equation}
For a given radius $r$ the 'impact' parameter $p$ can vary between 0 and $r$. Because the parameter $p$ cannot distinguish between $\mu >0$ and $\mu < 0$, the radiation intensity $I_{\nu}$ have to be separated into outward $I^{+}_{\nu}$ and inward $I^{-}_{\nu}$ directed intensity:
\begin{eqnarray}
  I^{+}_{\nu} &=& I_{\nu} \left(r,p\right) \hskip 1cm \mu \geq 0 \\
	I^{-}_{\nu} &=& I_{\nu} \left(r,p\right) \hskip 1cm \mu < 0  \textrm{.}
\end{eqnarray}

Now, in the new system of coordinates Eq.(\ref{sfer}) can be written as two separate equations
\begin{eqnarray}
	  \mu \frac{\partial I^{+}_{\nu}}{\partial r}& = & - \varrho \chi_{\nu} \left\{ I^{+}_{\nu}	-  S_{\nu}\left(r, p\right)	\right\} \label{one} \\
	  \mu \frac{\partial I^{-}_{\nu}}{\partial r} & = & \varrho \chi_{\nu} \left\{ I^{-}_{\nu}	-  S_{\nu}\left(r, p\right)\right\}	 \label{two} \textrm{,}
\end{eqnarray}
where $\mu = \sqrt{1-\frac{p^{2}}{r^{2}}}\geq 0$ and $S_{\nu}\left(r, p\right) = \eta_{\nu}\left(r, p\right)/\chi_{\nu}\left(r\right)$ is the source function.

If we introduce \cite{fe64} 
\begin{eqnarray}
	j_{\nu} \left(r,p\right) &=& \frac{1}{2} \left(   I_{\nu}^{+} \left(r,p\right) + I_{\nu}^{-} \left(r,p\right) \right) \label{fe1} \\
	h_{\nu} \left(r,p\right) &=& \frac{1}{2} \left(   I_{\nu}^{+} \left(r,p\right) - I_{\nu}^{-} \left(r,p\right) \right) \textrm{,} \label{fe2}
\end{eqnarray}
than Eqs.(\ref{one}) and (\ref{two}) will get the form
\begin{eqnarray}
	  \mu \frac{\partial j_{\nu}\left(r,p\right)}{\partial r}&=& 
	- \varrho \chi_{\nu} h_{\nu} \left(r,p\right) \label{t1} \\
	  \mu \frac{\partial h_{\nu} \left(r,p\right)}{\partial r}&=& -\varrho \chi_{\nu} \left\{ j_{\nu}\left(r,p\right)	-  S_{\nu}\left(r,p\right)	\right\} 
 \label{t2} \textrm{.} 
\end{eqnarray} 

To solve these equations we need also a boundary condition. We consider a spherical envelope with an inner boundary at radius $r=r_{min}$ and an outer boundary $r=r_{max}$. The cloud is immersed in an external time dependent, isotropic radiation field. We obtain the following boundary condition
\begin{eqnarray}
	h_{\nu} \left( r_{min}, p\right) &=& \alpha_{\nu} \left(p, t\right) 	\hskip 1 cm 0 \leq p \leq r_{min} \nonumber \\
	h_{\nu} \left(p,p\right) &=& 0 \hskip 1.5cm r_{min} \leq p \leq r_{max}  \label{w2} \\	
	j_{\nu} \left(r_{max}, p \right) - h_{\nu}\left(r_{max},p\right) &=& \beta_{\nu} \left(p, t\right) \hskip 1cm 0<p<r_{max}  \textrm{,} \nonumber
\end{eqnarray}
where $\alpha_{\nu}$ and $\beta_{\nu}$ describe the changes of the radiation at frequency $\nu$ impinging upon the inner and the outer boundary with the time $t$.

By the standard procedure one obtains from Eq.(\ref{sfer}) the zeroth and the first moment equations \cite{rad}
\begin{eqnarray}
	\frac{\partial \left( f_{\nu}\left(r\right) J_{\nu}\left(r\right) \right)}{\partial r} +\frac{3 f_{\nu}\left(r\right)-1}{r}J_{\nu} \left(r\right) + \varrho \chi_{\nu} H_{\nu} \left(r\right) &=& 0 \label{m0} \\
	\frac{\partial H_{\nu} \left(r\right)}{\partial r} + \frac{2 H_{\nu} \left(r\right)}{r} + \varrho \kappa_{\nu} J_{\nu} \left(r\right) - \varrho \eta_{\nu}^{t} &=& 0 \textrm{,} \label{m1}
\end{eqnarray}
where $f_{\nu}\left(r\right)=K_{\nu}\left(r\right)/J_{\nu}\left(r\right)$ is the Eddington factor. $J_{\nu}$, $H_{\nu}$, $K_{\nu}$ are the zeroth, first and second moment of the radiation field at frequency $\nu$:
\begin{eqnarray}
	J_{\nu} \left(r\right) &=&  \int_{0}^{1} j_{\nu} d\mu \\
	H_{\nu} \left(r\right) &=&  \int_{0}^{1} h_{\nu}  \mu d\mu \\
	K_{\nu} \left(r\right) &=&  \int_{0}^{1} j_{\nu} \mu^{2} d\mu \textrm{.}
\end{eqnarray}

We can cast Eq.(\ref{m0}) into a more convenient form, by introducing a sphericality factor $q_{\nu}$, defined in the following way:
\begin{equation}
	q_{\nu} \left(r\right) = \exp \left[  \int_{r_{c}}^{r} \left(3- \frac{1}{f_{\nu} \left(r'\right)} \right) \frac{dr'}{r'}\right] \textrm{,}
\end{equation}
where $r_{c}$ is the core radius, that is, the inner boundary of the medium. Using this factor and making some simplification we can rewrite Eqs.(\ref{m0}) and (\ref{m1}) as
\begin{eqnarray}
	\frac{\partial\left( f_{\nu} \left(r\right) q_{\nu} \left(r\right)J_{\nu} \left(r\right)\right)}{\partial r} &=& - \rho \chi_{\nu} q_{\nu}\left(r\right)  H_{\nu} \left(r\right)\label{ma0} \\
	\frac{\partial\left( H_{\nu} \left(r\right)r^{2}\right)}{\partial r} &=& r^{2} \rho \left( \eta_{\nu}^{t}  -\kappa_{\nu} J_{\nu} \left(r\right)\right) \textrm{.} \label{ma1}
\end{eqnarray} 

The corresponding boundary conditions for the moment equations (\ref{ma0}) and (\ref{ma1})) can be written as follows \cite{Yo80}
\begin{eqnarray}
	H_{\nu}\left(r_{min}\right) &=& \int_{0}^{1} \alpha_{\nu} \mu d\mu \label{w0} \\
	J_{\nu}\left(r_{max}\right) &=& \int_{0}^{1} \beta_{\nu} d \mu + \int^{1}_{0} h_{\nu} 
	\left(r_{max}, p\right) d\mu \label{w1} \textrm{.}
\end{eqnarray}

We see that these boundary condition can be determined only after Eqs.(\ref{t1}) and(\ref{t2}) have been solved for $j_{\nu}$ and $h_{\nu}$.

In the following sections we shall refer to Eqs.(\ref{t1}) and (\ref{t2}) along with the boundary conditions (\ref{w2}) as system I equations. The moment equations (\ref{ma0}) and (\ref{ma1}) along with the boundary condition (\ref{w0}), (\ref{w1}) will henceforth be referred as system II equations. 

Solution of the radiative energy transfer equations gives us information about the luminosity function $L\left(r\right)$:
\begin{equation}
	L\left(r\right) = 16 \pi^{2} r^{2} H\left(r\right)  \label{luminosity} \textrm{,}
\end{equation}
where
\begin{equation}
	H\left(r\right) = \int^{\infty}_{0} H_{\nu}\left(r\right) d \nu \textrm{.}
\end{equation}

With the given luminosity we can calculate radiative part of cooling function $\Lambda_{rad}$:
\begin{equation}
	\Lambda_{rad}\left(r\right)=-\frac{1}{4 \pi r^{2}} \frac{\partial L\left(r\right)}{\partial r} \label{co} \textrm{.}
\end{equation}
It can be used further in the hydrodynamical equations. 

\section{Chemical reactions and thermal effects}

In our calculations, we include all of the relevant thermal and chemical
processes in
the primordial gas. We have taken into account nine species:
H, H$^-$, H$^+$, He, He$^+$, He$^{++}$, H$_2$, H$_2^+$ and e$^-$.
Their abundance varies with time due to chemical reactions,
ionization and dissociation photoprocesses. The chemical reactions
include such processes as ionization of hydrogen and helium
by electrons, recombination of ions with electrons,
formation of negative hydrogen ions, formation of H$_2$
molecules, etc. A full list of the relevant chemical reactions and appropriate
formulae is given in \cite{Sta01}.

The time evolution of the number density of component $n_i$ is
described by the kinetic equation:

\begin{equation}
{dn_i \over dt} = \sum_{l=1}^9 \sum_{m=1}^9 a_{lmi} k_{lm} n_l n_m +
\sum_{j=1}^9 b_{ji} \kappa_j n_j .
\label{chemia}
\end{equation}

\ni The first component on the right-hand side of this equation describes
the chemical reactions, and the other one accounts for photoionization and
photodissociation processes. Coefficients
$k_{lm}$ are reaction rates, quantities $\kappa_n$ are photoionization
or photodissociation
rates, and $a_{lmi}$ and $b_{ji}$ are numbers equal to 0, $\pm 1$ or $\pm 2$
depending on the reaction.
All reaction rates, as well as photoionization and photodissociation rates
are given in \cite{Sta01}.

The cooling (or heating) function $\Lambda$ includes effects of
collisional ionization of H, He and He$^+$, recombination to H, He and He$^+$,
collisional excitation of H and He$^+$, Bremsstrahlung, Compton cooling
and cooling by de-excitation of H$_2$ molecules. 
The heating/cooling rates are given in \cite{Sta01}.

\section{Numerical strategy and initial conditions}

In the simulations we used the code described in \cite{Sta01}, based on
those presented by \cite{Tho95} and \cite{Hai96}.
This is a standard, one-dimensional, second-order
accurate, Lagrangian finite-difference scheme, slightly modified to our 
purposes. Our code supports flat and non-flat CDM and $\Lambda$CDM models 
as well as Milgrom's Modified Newtonian Dynamics. In the second 
case it was necessary to do significant changes in the initial conditions. 
We starte our calculations at the end of the
radiation-dominated era instead of $z=500$. For $\Omega_b=\Omega_M=0.02/h^2$, 
$z_{eq}=485$ as given by the formula provided by \cite{HE98},
$z_{eq}=2.50 \times 10^4 \Omega_0 h^2 \Theta_{2.7}^{-4}$,
where $\Theta_{2.7}=T_\gamma/2.7$ K, assuming $h=0.72$ and
$T_\gamma$=2.7277 K. We assume that in MOND, like in the standard
cosmology, initial overdensities may grow only in the matter-dominated
era. In both cases we use our own code to calculate the initial chemical 
composition and initial gas temperature. Initial overdensities may 
be calculated from the matter power spectrum which may be obtained e.g. 
using the {\sc CMBFAST} program by \cite{Sel96}. 

We apply the initial density profiles in the form of a single spherical
Fourier mode, also used by \cite{Hai96}

\begin{equation} \varrho_b(r)=\Omega_b \varrho_c (1+\delta {\sin kr \over
kr})
,
\label{psinus} \end{equation}

\ni where $\varrho_c$ is the critical density of the Universe,
$\varrho_c=3H^2/8\pi G$, with $H$ being the actual value of the Hubble
parameter.

For this profile, we can distinguish two radius values, $R_0$ and $R_z$, 
which correspond to the first zero and the first minimum of
the function $\sin (kr)/kr$, respectively.
Inside the sphere of radius $R_0=\pi/k$ which contains mass
$M_0$, the local density contrast is positive. The
mass $M_0$ and the radius $R_0$ will be referred to as the cloud
mass and the cloud radius, respectively. The local
density contrast is negative for $R_z>r>R_0$, with average
density contrast vanishing for the sphere of radius $R_z=4.49341/k$ and
mass $M_z$. The shell of radius $R_z$ will expand
together with the Hubble flow, not undergoing any additional deceleration.
This is why this profile is very convenient in numerical simulations.
It eliminates the numerical edge effects and the mentioned shell simply
follows the Hubble expansion of the universe. Thus, it can be regarded as the
perturbation boundary, with mass $M_z$ referred to as the bound mass.

It is worth to note that for radii not greater than $3/4 R_0$ this profile
is very similar to the Gaussian profile

\begin{equation} \varrho_i(r)=\Omega_i \varrho_c \left[ 1+\delta_i \exp \left(
{-r^2 \over 2R_{\rm f}^2} \right) \right] \end{equation}

\ni with $R_{\rm f}=1/2 R_0$.

As the initial velocity, we use the Hubble velocity:

\begin{equation}v(r)=Hr . \end{equation}

Our numerical computational procedure to trace the dynamical evolution of the primordial cloud under the UV background is following:
\begin{itemize}
	\item We solve the hydrodynamic equations of motion along with equations for energy conservation, ionization, and dissociation of molecular and atomic species. From the solution of these equations we can calculate the properties and distribution of the absorbing components $\kappa_{\nu}\left(r\right)$, $\sigma_{\nu}\left(r\right)$ and the thermal emissivity $\eta^{t}_{\nu}\left(r\right)$. It will give us the initial value of the source function $S_{\nu}$.
	\item With the initial value of $S_{\nu}$, we solve the system I equations for the geometrical factors $f_{\nu}$ and $q_{\nu}$.
  \item We solve the system II equations for $J_{\nu}$ and $H_{\nu}$.
  \item The source function is independent from the radiation field ($J_{\nu}$) only when there is no light scattering. So, in general we have to update $S_{\nu}$ and solve the system I once more.
  \item Iterative procedure between system I and II is continued until convergence.
  \item We calculate luminosity $L\left(r\right)$ from Eq.(\ref{luminosity}) and than cooling function.
  \item We update abundance of different species and number densities of each atomic and molecular state.
  \item We repeat all of the above steps.

\end{itemize}

\section{Results}

Detailed results of our simulations will be described in a paper, which is in preparation.


\begin{thebibliography}{99}

\bibitem{Do02} Doroshkevich A. G., Naselsky P. D., 2002, astro-ph/0201212

\bibitem{Sh04} Shchekinov Y. A., Vasiliev E. O., 2004, A\&A {419}, 19

\bibitem{Ms82} Shull J. M., Beckwith S., 1982, Ann. Rev. Astron. Astrophys. {20}, 163

\bibitem{Ce03} Cen R., 2003, ApJ {591}, 12 

\bibitem{Ta98} Tajiri Y., Umemura M., 1998, ApJ {502}, 59

\bibitem{Ki00} Kitayama T., Ikeuchi S., 2000, ApJ {529}, 615

\bibitem{Ki00b} Kitayama T., Tajiri Y., Umemura M., Susa H., Ikeuchi S., 2000, MNRAS {315}, L1

\bibitem{Ki00c} Kitayama T., Susa H., Umemura M., Ikeuchi S., 2001, MNRAS {326}, 1353

\bibitem{Om01} Omukai K., 2001, ApJ {546}, 635

\bibitem{Dr96} Draine B. T., Bertoldi F., 1996, ApJ {468}, 269

\bibitem{Mil83} Milgrom M., 1983, ApJ, {270}, 371

\bibitem{Sta05} Stachniewicz S., Kutschera M., 2005, MNRAS {362}, 89

\bibitem{hu71} Hummer D. G., Rybicki G. B., 1971, MNRAS {152}, 1

\bibitem{fe64} Feautrier P., 1964, C. r. hebd. Seance Acad. Sci. Paris 258, 3189

\bibitem{rad} Mihalas D., Mihalas B. W., 1984, 'Foundations of Radiation Hydrodynamics', (New York: Oxford Univ. Press)

\bibitem{Yo80} York H. W., 1980, A\&A {86}, 286

\bibitem{Sta01} Stachniewicz S., Kutschera M., 2001, Acta Phys. Pol. B {32}, 227

\bibitem{Tho95} Thoul A. A., Weinberg D. H., 1995, ApJ, {442}, 480

\bibitem{Hai96} Haiman Z., Thoul A. A., Loeb A., 1996, ApJ, {464}, 523

\bibitem{HE98} Hu W., Eisenstein D. J, 1998, ApJ, {498}, 497

\bibitem{Sel96} Seljak U., Zaldarriaga M., 1996, ApJ {469}, 437

\end{thebibliography}
\end{document}